\documentclass[prd,twocolumn,showpacs]{revtex4}
\usepackage[dvips]{graphicx}
\usepackage{color}
\usepackage{latexsym}
\usepackage{amsmath}
\usepackage{amssymb}
\newcommand{\rd}{\mathrm{d}}
\newcommand{\C}{\mathcal{C}}

\begin{document}
%
%
%
%
\preprint{LAUR 03-1817}
\title[O(N) model]
   {Real-time dynamics of the O(N) model in 1+1 dimensions}
\author{Bogdan Mihaila}
\email{bogdan.mihaila@unh.edu}
\affiliation{Theoretical Division,
   Los Alamos National Laboratory,
   Los Alamos, NM 87545}
\date{\today}
\begin{abstract}
We study the non-equilibrium dynamics of the O(N) model in
classical and quantum field theory in 1+1 dimensions, for $N > 1$.
We compare numerical results obtained using the Hartree
approximation and two next to leading order approximations, the
bare vertex approximation and the 2PI-1/N expansion. The later
approximations differ through terms of order $g^2$, where $g$ is
the scaled coupling constant, $g=\lambda/N$. In this paper we
investigate the statement regarding the convergence with respect
to $g$. We find that the differences between these two
approximation schemes diminish for larger values of $N$, when
$\lambda$ is fixed.
\end{abstract}
\maketitle
%
%

\section{Introduction}

Initial value problems in quantum field theory represent a key
step in the pursuit of first principle understanding of physics of
the early universe and dynamics of phase transitions and particle
production following a relativistic heavy-ion collision. In order
to achieve quantitative comparisons with experimental data, it is
absolutely necessary to go beyond existing mean field theoretical
models, and implement approximations which account for higher
order corrections. Thus we will not only have a reliable picture
of the phase diagram of quantum field theories, but the model will
include important physics such as a mechanism of allowing an out
of equilibrium system to be driven back to equilibrium, mechanism
which is absent in leading order. Much progress has been achieved
in recent years (see Ref.{r:BS03}, and references therein.)

We have recently~\cite{r:CDMq02} presented a first study of the
dynamics of a single component ``explicitly broken symmetry''
$\lambda \, \phi^4$ field theory in 1+1 dimensions, using a
Schwinger-Dyson (SD) equation truncation scheme based on ignoring
vertex corrections we call the bare vertex approximation
(BVA)~\cite{r:MCD01}. We have compared the BVA prediction for the
time evolution of the system with results obtained using two other
approximations: the Hartree approximation which represents a
leading order (LO) result in our three-point vertex function
truncation scheme, and the 2PI-1/N expansion~\cite{r:AABBS}, which
is a related next to leading order (NLO) approximation scheme. The
2PI-1/N expansion can be formally obtained from the BVA by
dropping a graph from the two-particle irreducible (2PI) effective
action~\cite{r:CJT, r:LW, r:Baym62}.

Two important aspects in connection with this previous study
deserve attention: Firstly, we have observed that the BVA
approximation of the O(1) model in 1+1 dimensions offers a unique
toy problem which allows for studying the dynamics of phase
transitions. It is well known that in 1+1 dimensions there is no
phase transition in the O(1) model except at zero
temperature~\cite{r:Griffiths}. Nevertheless, we used this model
to demonstrate some of the features expected to be true in 3+1
dimensions such as the restoration of symmetry breakdown at high
temperatures and equilibration of correlation functions. For this
model, we have found evidence that the order of the phase
transition predicted by the static phase diagram in the Hartree
approximation, is relaxed from a first order to second order phase
transition in the BVA. Moreover, unlike in the case of the Hartree
approximation where the order parameter oscillates about the
Hartree minimum without equilibration, the BVA results track the
Hartree curve, except with damping: the average fields in the BVA
equilibrate for arbitrary initial conditions. The 2PI-1/N
expansion goes to a zero value of the order parameter, in
agreement with the exact result of no phase transition for finite
temperature.

Secondly, we have shown that the disagreement between the 2PI-1/N
and BVA predictions for the order parameter is more pronounced at
larger values of the coupling constant $\lambda$, while
predictions for the effective mass 
are similar for both approximations, in agreement with
expectations based on our experience with the classical limit of
these approximations~\cite{r:CDM02}. We remind the reader that in
the classical field theory case in 1+1 dimensions we were able to
compare these approximations with exact Monte Carlo
simulations~\cite{r:MD02}. Note also that in the classical case
the quality of the agreement between approximation and exact
results is not sensitive to the coupling constant value, since the
coupling can be scaled out in the classical limit. This scaling is
\emph{not} possible in the quantum case. Nevertheless, direct
comparisons with exact results are currently possible only in the
classical case, and the differences between the BVA and the exact
results were said to be due to the fact that three-point vertex
corrections become important at intermediate time scales, and they
affect the evolution of the two-point function more than the time
evolution of the order parameter.

The study of the O(1) model in 1+1 dimensions carried out
in~\cite{r:CDMq02} was both exciting and worrisome: exciting
because the model provided a testing ground where features of the
theory expected to be true in 3+1 dimensions could be studied;
worrisome because the huge discrepancy between the 2PI-1/N and the
BVA results made it look hopeless to think that one can accurately
predict the time evolution of the system. It is imperative to
achieve convergence with the truncation order of the SD hierarchy
of equations, in order to allow for reliable predictions of the
dynamics of the system, and the results for the O(1) model in 1+1
dimensions imply that the difference between the 2PI-1/N and BVA
approximations is considerable. Since the formal difference is
related to the additional graph in the 2PI effective action
considered in the BVA, the importance of this graph appeared to be
considerable. Since the physical system we are interested in is in
fact the O(4) linear sigma model in 3+1 dimensions, one can only
hope that by increasing the number of dimensions and/or the number
of fields, the agreement between these methods will get better. In
this paper we present an immediate extension of the work discussed
in~\cite{r:CDMq02}, by generalizing the calculations from the O(1)
to the O(N) model, in 1+1 dimensions.


\section{BVA equations}
\label{sec:bva}

In this section we briefly review the basics of the BVA formalism
(for full details, see Ref.~\cite{r:MCD01}): We start with the
Lagrangian for the O(N) symmetry model
\begin{align}
   \mathcal{L}[\phi,\partial_\mu \phi]
    = &
   \frac{1}{2} \,
      \Bigl [
         \partial_\mu \phi_i(x) \, \partial^\mu \phi_i(x)
         +
         \mu^2 \, \phi_i^2(x)
      \Bigr ]
   \notag \\ &
   -
   \frac{\lambda}{8} \, [ \phi_i^2(x) ]^2
   -
   \frac{\mu^4 }{2 \lambda} \>.
\end{align}
Here the Einstein summation convention for repeated indices is
implied. We introduce a composite field $\chi(x)$ by adding to the
Lagrangian a term:
\begin{equation}
   + \
   \frac{1}{2 \lambda} \,
   \left \{
      \chi(x) - \frac{\lambda}{2} \,
      \Bigl [ \, \phi_i^2(x) - 
                 2 \frac{\mu^2}{\lambda} \, \Bigr ]
   \right \}^2
   \>.
\end{equation}
This gives a Lagrangian of the form
\begin{align}
   \mathcal{L}[\phi, \chi, \partial_\mu \phi]
   = &
   \frac{1}{2} \,
      \Bigl [
         \partial_\mu \phi_i(x) \, \partial^\mu \phi_i(x)
         - \chi(x) \, \phi_i^2(x)
      \Bigr ]
   \notag \\ &
   +
   \frac{ \mu^2 \, \chi(x) }{\lambda}
   +
   \frac{ \chi^2(x) }{ 2 \, \lambda }  \>,
\end{align}
which leads to the classical equations of motion
\begin{gather}
   \Bigl  [ \Box +  \chi(x) \Bigr ] \, \phi_i(x)
   = 0 \>,
\end{gather}
and the constraint (``gap'') equation
\begin{gather}
   \chi(x)
   =
   - \mu^2
   +
   \frac{\lambda}{2} \, \sum_i \, \phi_i^2(x)
   \>.
\end{gather}
It is convenient now to introduce the super-fields and
super-currents, $\phi_\alpha(x)=\{
\chi(x)$,$\phi_1(x)$,$\phi_2(x)$,$\ldots$,$\phi_N(x)\}$, and
$j_\alpha(x)=\{ J(x)$,$j_1(x)$,$j_2(x)$,$\ldots$,$j_N(x)\}$, with
$\alpha = 0, 1, \ldots, N$. Then, the BVA equations can be
obtained from the effective action
\begin{equation}
   \Gamma[\phi,G]
   =
   S_{\text{cl}}[\phi]
   +
   \frac{i}{2} {\rm Tr} \ln [ G^{-1} ]
   +
   \frac{i}{2} {\rm Tr} [ G_0^{-1} G ]
   +
   \Gamma_2[G] \>,
   \label{e:BVAaction}
\end{equation}
where $ \Gamma_2[G]$ is the Cornwall-Jackiw-Tomboulis 
generating functional of the two-particle irreducible (2-PI)
graphs. Here $G$ denotes the Green function matrix
\begin{align}
   G_{\alpha \beta}[j](x,x')
   &=
   \frac{\delta \phi_{\alpha}(x)}{\delta j_{\beta}(x')}
   =
   \begin{pmatrix}
      D(x,x')         & K_{j}(x,x') \\
      \bar{K}_i(x,x') & G_{i j}(x,x')
   \end{pmatrix}  \>,
\end{align}
and $S_{\text{cl}}[\phi]$ is the classical action in Minkowski
space
\begin{align}
   S_{\text{cl}}[\phi]
   =
   \int d^2x \,
   \Bigl \{ &
   - \frac{1}{2}
   \phi_i(x) \,
   [ \, \Box + \chi(x) \, ] \, \phi_i(x)
   \notag \\ &
   +
   \frac{\chi^2(x)}{2 \lambda}
   +
   \frac{\mu^2}{\lambda} \chi(x) \,
   \Bigr \} \>.
\end{align}
In the BVA approximation, we keep in $\Gamma_2[G]$ only the terms
\begin{align}
   \Gamma_2[G]
   =
   - \frac{1}{4}
   \iint \rd x \, \rd y
   \Bigl [ &
      G_{ij}(x,y) G_{ji}(y,x) D(x,y)
   \\ \notag &
      +
      2 \bar{K}_i(x,y) K_j(x,y) G_{ij}(x,y)
   \Bigr ]
   \>.
   \label{e:Gamma2}
\end{align}
This is equivalent to a truncation of the SD tower of equations,
where one approximates the exact three-point vertex function
equation
\begin{align}
   \Gamma_{\alpha \beta \gamma}[\phi](x_1,x_2,x_3)
   = &
   f_{\alpha \beta \gamma} \,
   \delta_{\C}(x_1,x_2) \, \delta_{\C}(x_1,x_3)
   \notag \\ &
   +
   \frac{\delta \, \Sigma_{\alpha \beta}[\phi](x_1,x_2)}
        {\delta \phi_\gamma(x_3)}  \>.
\end{align}
Here $f_{i j 0} = f_{0 i j} = f_{i 0 j} = \delta_{ij}$, and $f$ is
zero otherwise, and $\Sigma_{\alpha \beta}[\phi](x,x')$ denote the
elements of the self-energy matrix. Then, the Hartree
approximation corresponds to the case when one ignores the
three-point vertex corrections altogether, while in the BVA one
keeps only the contact term. Finally, the 2PI-1/N approximation is
obtained from the BVA approximation by dropping the second term in
$\Gamma_2[G]$, see Eq.\eqref{e:Gamma2}.


The following equations of motion
\begin{gather}
   \Bigl [ \Box + \chi(x) \Bigr ] \, \phi_i(x)
   + K_i(x,x) / i= 0 \>,
   \\
   \chi(x) = - \mu^2
   + \frac{g}{2} \sum_i
   \Bigl [ \phi_i^2(x) + G_{ii}(x,x)/i \Bigr ] \>,
\end{gather}
must be solved self-consistently with the Green function equations
\begin{equation}
   G_{\alpha \beta}^{-1}(x,x')
   =
   G_{0\, \alpha \beta}^{-1}(x,x')
   + \Sigma_{\alpha \beta}(x,x') \>, 
\end{equation}
which are subject to the condition
\begin{equation}
   \int_{\C} \mathrm{d}x'' \
       G_{\alpha \beta}^{-1}(x,x'') \,
       G_{\beta \gamma}(x'',x')
   = \delta_{\alpha \gamma} \delta_{\C}(x,x') \>.
\end{equation}
Here $g=\lambda/N$, and this indicates that the expansion
parameter in this formalism is $g$, rather then the usual $1/N$
corresponding to the familiar large-N expansion. The integrals and
delta functions $\delta_{\C}(x,x')$ are defined on the closed time
path (CTP) contour, which incorporates the initial value boundary
condition \cite{r:Schwinger,r:Keldish,r:MahanI,r:MahanII}. The
elements of the Green function $G_{0\, \alpha \beta}^{-1}(x,x')$
are
\begin{align}
   G_{0\, 00}^{-1} &\equiv D_0^{-1}(x,x') = - \frac{1}{g} \, \delta_{\C}(x,x') \>,
   \label{eq:G0} \\
   G_{0\, ij}^{-1} &\equiv G_0^{-1}(x,x') = \Bigl [ \Box + \chi(t) \Bigr ] \, \delta_{ij} \, \delta_{\C}(x,x') \>,
   \notag \\
   G_{0\, 0j}^{-1} &\equiv K_{0\,j}^{-1}(x,x') = \phi_j(x) \, \delta_{\C}(x,x')\>,
   \notag \\
   G_{0\, i0}^{-1} &\equiv \bar K_{0\,i}^{-1}(x,x') = K_{0\,i}^{-1}(x',x) \>.
   \notag
\end{align}
For the BVA, the self-energies $\Sigma^{\text{BVA}}_{\alpha
\beta}(x,x')$ are given as
\begin{align}
   \Pi(x,x')
   &= \frac{i}{2} \, G_{ij}(x,x') \, G_{ij}(x,x') \>,
   \label{eq:SigmasBVA} \\
   \Omega_i(x,x')
   &= i \, \bar K_j(x,x') \, G_{ji}(x,x') \>,
   \notag \\
   \bar \Omega_i(x,x')
   &= i \, G_{ij}(x,x') \, K_j(x,x') \>,
   \notag \\
   \Sigma_{ij}(x,x')
   &= i \, \Bigl [ G_{ij}(x,x') \, D(x,x') + \bar K_i(x,x') \, K_j(x,x') \Bigr ] \>.
   \notag
\end{align}
The equations for the 2PI-1/N approximation can be formally
obtained by setting $\Omega_i(x,x')=\bar \Omega_j(x',x)=0$, and
dropping the  $\bar K_i K_j$ term in the definition of
$\Sigma_{ij}(x,x')$.

The BVA is an energy-conserving approximation, where the average
energy is given as
\begin{align}
   E = &
   \frac{\mu^2}{\lambda} \langle \chi \rangle
   - \frac{1}{2 \lambda} \langle \chi^2 \rangle
   + \frac{1}{2} \Bigl \{
   \langle \bigl [ \partial_t \phi \bigr ]^2 \rangle
   + \langle \bigl [ \partial_x \phi \bigr ]^2 \rangle
   + \langle \chi \, \phi^2 \rangle \Bigr \}
   \>,
\end{align}
with the expectation values
\begin{align}
   \langle \chi^2 \rangle
   = &
   \chi^2(x) + D(x,x)/i
   \>,
\\
   \langle \phi_i^2 \rangle
   = &
   \phi_i^2(x) + G_{ii}(x,x)/i
   \>,
\\
   \langle \chi \, \phi^2 \rangle
   = &
   \chi(x) \sum_i \langle \phi_i^2 \rangle
   \ + \ 2 \, \phi_i(x) K_i(x,x)/i
   \\ \notag &
   + \frac{2}{3}
   \int_{\C} \mathrm{d}x' \Bigl [
       \Pi(x,x') D(x',x) + \bar \Omega_i(x,x') \bar K_i(x',x)
   \\ \notag & \qquad
       + \Omega_i(x,x') K_i(x',x) + \Sigma_{ij}(x,x') G_{ji}(x',x)
   \Bigr ]
   \>.
\end{align}
Since the energy calculation involves only second order
derivatives of the effective action, it results that any
truncation scheme done at the level of the three-point vertex
function or beyond, and which produces a self-energy matrix
obeying the appropriate symmetries, will be an energy-conserving
approximation.

In order to investigate the convergence of the proposed
approximations with respect to $g$, we consider two initial
scenarios which, from the point of view of computational storage
requirements, differ only minimally from the O(1) calculations we
have previously performed. We summarize next the explicit form of
the equations we have to solve, and propose a practical way of
obtaining the numerical solution.

\subsection{One-field scenario}

Firstly, in what we call the ``one-field scenario", we study the
case when only one of the initial fields, denoted $\phi_1$, is
nonzero, i.e. $\phi_2(0) = \cdots = \phi_N(0) = 0$, and $\dot
\phi_2(0) = \cdots = \dot \phi_N(0) = 0$. In this case, we can
show that the Green function matrix remains diagonal at all times,
and all diagonal matrix elements other than $G_{11}$ are
identical. Therefore, in the one-field scenario the Green function
and self-energy matrices can be written as
\begin{align}
   G_{\alpha \beta}^{-1}
   =
   \begin{pmatrix}
      D_0^{-1} & K_0^{-1} & 0 & \cdots & 0 \\
      \bar K_0^{-1} & G_0^{-1} & 0 & \cdots & 0 \\
      0 & 0 & G_0^{-1} & \cdots & 0 \\
      \cdots & \cdots & \cdots & \cdots & \cdots \\
      0 & 0 & 0 & \cdots & G_0^{-1}
   \end{pmatrix}
   \>,
   \\
   \Sigma_{\alpha \beta}
   =
   \begin{pmatrix}
      \Pi & \Omega & 0 & \cdots & 0 \\
      \bar \Omega & \Sigma_{11} & 0 & \cdots & 0 \\
      0 & 0 & \Sigma_{22} & \cdots & 0 \\
      \cdots & \cdots & \cdots & \cdots & \cdots \\
      0 & 0 & 0 & \cdots & \Sigma_{22}
   \end{pmatrix}
   \>,
   \\
   G_{\alpha \beta}
   =
   \begin{pmatrix}
      D & K_1 & 0 & \cdots & 0 \\
      \bar K_1 & G_{11} & 0 & \cdots & 0 \\
      0 & 0 & G_{22} & \cdots & 0 \\
      \cdots & \cdots & \cdots & \cdots & \cdots \\
      0 & 0 & 0 & \cdots & G_{22}
   \end{pmatrix}
   \>.
\end{align}
We find it convenient to introduce the additional Green functions
\begin{subequations}
\label{eq:one_field_notations}
\begin{align}
   D_2^{-1}(x,x') & = D_0^{-1}(x,x') + \Pi(x,x')\>, \\
   G_1^{-1}(x,x') & = G_0^{-1}(x,x') + \Sigma_{11}(x,x') \>, \\
   G_2^{-1}(x,x') & = G_0^{-1}(x,x') + \Sigma_{22}(x,x') \>, \\
   \Xi_{0}(x,x')  & = K_{0}^{-1}(x,x') + \Omega(x,x') \>.
\end{align}
\end{subequations}
With these notations, the Green function equations are
\begin{align}
   1 & = D_2^{-1} \, D + \Xi_0 \, \bar K \>,
\label{eq:one_field_eqs} \\
   0 & = D_2^{-1} \, K + \Xi_0 \, G_{11} \>,
   \notag \\
   0 & = \bar \Xi_0 \, D + G_1^{-1} \, \bar K \>,
   \notag \\
   1 & = \bar \Xi_0 \, K + G_1^{-1} \, G_{11} \>,
   \notag \\
   1 & = G_2^{-1} \, G_{22} \>.
   \notag
\end{align}
The numerical solution is achieved by iterating the following
system of equations
\begin{align}
   K &= - \, D_2 \, \Xi_0 \, G_{11} \>,
\label{eq:one_field_solition} \\
   \bar K &= - \, G_1 \, \bar \Xi_0 \, D \>,
   \notag \\
   D &= D_2 + D_2 \, \Bigl [ \Xi_0 \, G_1 \, \bar \Xi_0 \Bigr ] \, D \>,
   \notag \\
   G_{11} &= G_1 + G_1 \, \Bigl [ \bar \Xi_0 \, D_2 \, \Xi_0 \Bigr ] \, G_{11} \>,
   \notag \\
   G_{22} &= G_2 \>.
   \notag
\end{align}

\subsection{N-copies scenario}

Secondly, we study the case when initially we have~N identical
copies of the field, i.e. $\phi_i(0) = \phi_0$ and $\dot \phi_i(0)
= \pi_0$, for $i =1 \ldots N$. In this case we can still choose
the Green function to be diagonal at $t=0$, but this property is
lost as we propagate the equations of motion. However, the Green
function matrix in the ``N-copies scenario" remains particularly
simple, as the diagonal and off-diagonal matrix elements are
separately identical.

Once again, we start with the Green function and self-energy
matrices, which in this case have the form
\begin{align}
   G_{\alpha \beta}^{-1}
   =
   \begin{pmatrix}
      D_0^{-1} & K_0^{-1} & K_0^{-1} & \cdots & K_0^{-1} \\
      \bar K_0^{-1} & G_0^{-1} & 0 & \cdots & 0 \\
      \bar K_0^{-1} & 0 & G_0^{-1} & \cdots & 0 \\
      \cdots & \cdots & \cdots & \cdots & \cdots \\
      \bar K_0^{-1} & 0 & 0 & \cdots & G_0^{-1}
   \end{pmatrix}
   \>,
   \\
   \Sigma_{\alpha \beta}
   =
   \begin{pmatrix}
      \Pi & \Omega & \Omega & \cdots & \Omega \\
      \bar \Omega & \Sigma_{11} & \Sigma_{12} & \cdots & \Sigma_{12} \\
      \bar \Omega & \Sigma_{21} & \Sigma_{11} & \cdots & \Sigma_{12} \\
      \cdots & \cdots & \cdots & \cdots & \cdots \\
      \bar \Omega & \Sigma_{21} & \Sigma_{21} & \cdots & \Sigma_{11}
   \end{pmatrix}
   \>,
   \\
   G_{\alpha \beta}
   =
   \begin{pmatrix}
      D & K & K & \cdots & K \\
      \bar K & G_{11} & G_{12} & \cdots & G_{12} \\
      \bar K & G_{21} & G_{11} & \cdots & G_{12} \\
      \cdots & \cdots & \cdots & \cdots & \cdots \\
      \bar K & G_{21} & G_{21} & \cdots & G_{11}
   \end{pmatrix}
   \>.
\end{align}
Correspondingly, the Green functions equations we have to solve
are
\begin{align}
   1 & = D_2^{-1} \, D + {\mathrm N} \, \Xi_0 \, \bar K \>,
\label{eq:Ncopies_eqs} \\
   0 & = D_2^{-1} \, K + \Xi_0 \, \Bigl [ G_{11} + ( {\mathrm N} - 1 ) G_{12} \Bigr ] \>,
   \notag \\
   0 & = \bar \Xi_0 \, D + \Bigl [ G_1^{-1} + ( {\mathrm N} - 1 ) \Sigma_{12} \Bigr ] \bar K \>,
   \notag \\
   1 & = \bar \Xi_0 \, K + G_1^{-1} \, G_{11} + ( {\mathrm N} - 1 ) \Sigma_{12} \, G_{12} \>,
   \notag \\
   0 & = \bar \Xi_0 \, K + G_1^{-1} \, G_{12} + \Sigma_{12} \, G_{11}
         + ( {\mathrm N} - 2 ) \Sigma_{12} \, G_{12} \>.
   \notag
\end{align}
In our strategy, it is convenient to introduce the additional set
of Green functions
\begin{subequations}
\label{eq:Ncopies_notations}
\begin{align}
   G_a^{-1} & = G_1^{-1} - \bar \Xi_0 \, D_2 \, \Xi_0 \>, \\
   G_b^{-1} & = G_1^{-1} - \Sigma_{12} \>, \\
   G_c^{-1} & = G_1^{-1} + ( {\mathrm N} - 1 ) \Sigma_{12} \>.
\end{align}
\end{subequations}
Then, we obtain the desired Green functions by solving the
equations
\begin{align}
   K & = - \, D_2 \, \Xi_0 \, \Bigl [ G_{11} + ( {\mathrm N} - 1 ) G_{12} \Bigr ] \>,
\label{eq:Ncopies_solutions} \\
   \bar K & = - \, G_c \, \bar \Xi_0 \, D \>,
   \notag \\
   D & = D_2 + {\mathrm N} D_2 \, \Bigl [ \Xi_0 \, G_c \, \bar \Xi_0 \Bigr ] \, D \>,
   \notag \\
   G_{11} & = G_a - ( {\mathrm N} - 1 ) G_a \,
              \Bigl [ \Sigma_{12} - \bar \Xi_0 \, D_2 \, \Xi_0 \Bigr ] \, G_{12} \>,
   \notag \\
   G_{12} & = G_{11} - G_b \>.
   \notag
\end{align}

\section{Results}
\label{sec:results}

We study the O(N) model, in 1+1 dimensions. We set $\lambda=7.3$
since this particular value of $\lambda$ was used in previous
studies of the dynamics of disoriented chiral condensates (DCC) in
$3+1$ dimensions in the leading order in large-N
approximation~\cite{r:KDC96,r:KCMPKnp95}. We remind the reader
that $g=\lambda/N$, and this is the actual expansion parameter in
this approximation.

We assume here that the initial state is described by a Gaussian
density matrix peaked around some non-zero value of $\langle
\phi(0) \rangle$, and characterized by a single particle
Bose-Einstein distribution function at a given temperature. The
details of this choice of initial conditions has been discussed in
detail in Ref.~\cite{r:CDMq02}. In this study, we choose a very
low initial temperature of $T_0=0.1$ in order to emphasize quantum
effects in the dynamics. We set the initial values of the
non-vanishing initial fields to $\phi_0=\phi(0)/N = 1$, and
$\pi_0=\dot \phi(0)=0$.

The numerical procedure for solving the BVA equations is described
in detail in Refs.~\cite{r:MM02,r:MS02}. The unknown functions are
expanded out in terms of Chebyshev polynomials, on a nonuniform
grid. The algorithm follows a multi-step approach, and the
algorithm possess spectral convergence. We achieve convergence of
the numerical results with a relatively small number of grid
points, only 32 and 128 points, for the time step and the momentum
domain discretization, respectively. The self-energy matrix
elements are calculated using standard fast-Fourier transform
algorithms on a uniform 1024 points grid. We employ a cubic-spline
interpolation technique to perform the necessary transformation
between the two momentum grids. The mass renormalization necessary
for the quantum field theory calculations has been discussed
previously~\cite{r:CDMq02}. A momentum cutoff, $\Lambda=3\pi$, was
chosen for the purpose of the present calculation.

We study the time evolution of the one-point functions: the order
parameter $\langle \phi(t) \rangle$, and the auxiliary field
$\langle \chi(t) \rangle$. We compare the Hartree, 2PI-1/N
expansion and the BVA approximation. Results are presented for
both the classical (CFT) and the quantum field theory (QFT) case.
In keeping with our earlier work, the potential is a double-well
potential for the quantum case, but single-well in the classical
case.

\begin{figure}[h!]
   \includegraphics[height=0.26\textheight]{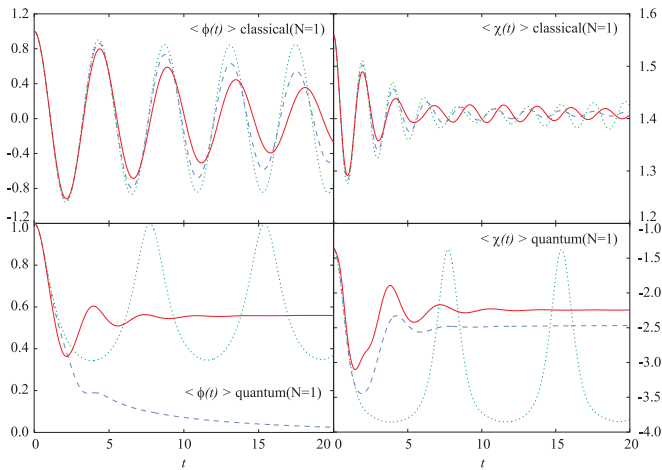}
   \caption{\label{fig:fig_N1} N=1;
   legend: \textcolor{red}{BVA} (solid) ,
   \textcolor{blue}{2PI-1N} (dashed), and
   \textcolor{green}{Hartree} (dotted)}
\end{figure}

We start by reviewing the results for the O(1) model. In
Fig.~\ref{fig:fig_N1} we present the time evolution of $\langle
\phi(t) \rangle$ and $\langle \chi(t) \rangle$ for CFT and QFT.

\begin{figure}[h!]
   \includegraphics[height=0.26\textheight]{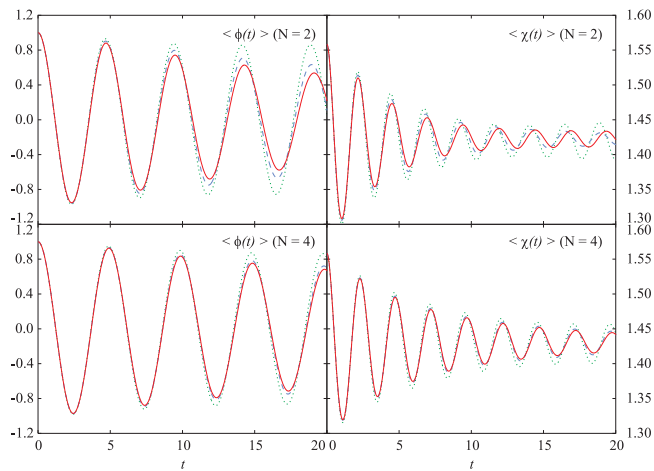}
   \caption{\label{fig:one_field_cl} One-field scenario: classical field
   theory.
   }
\end{figure}

\begin{figure}[h!]
   \includegraphics[height=0.26\textheight]{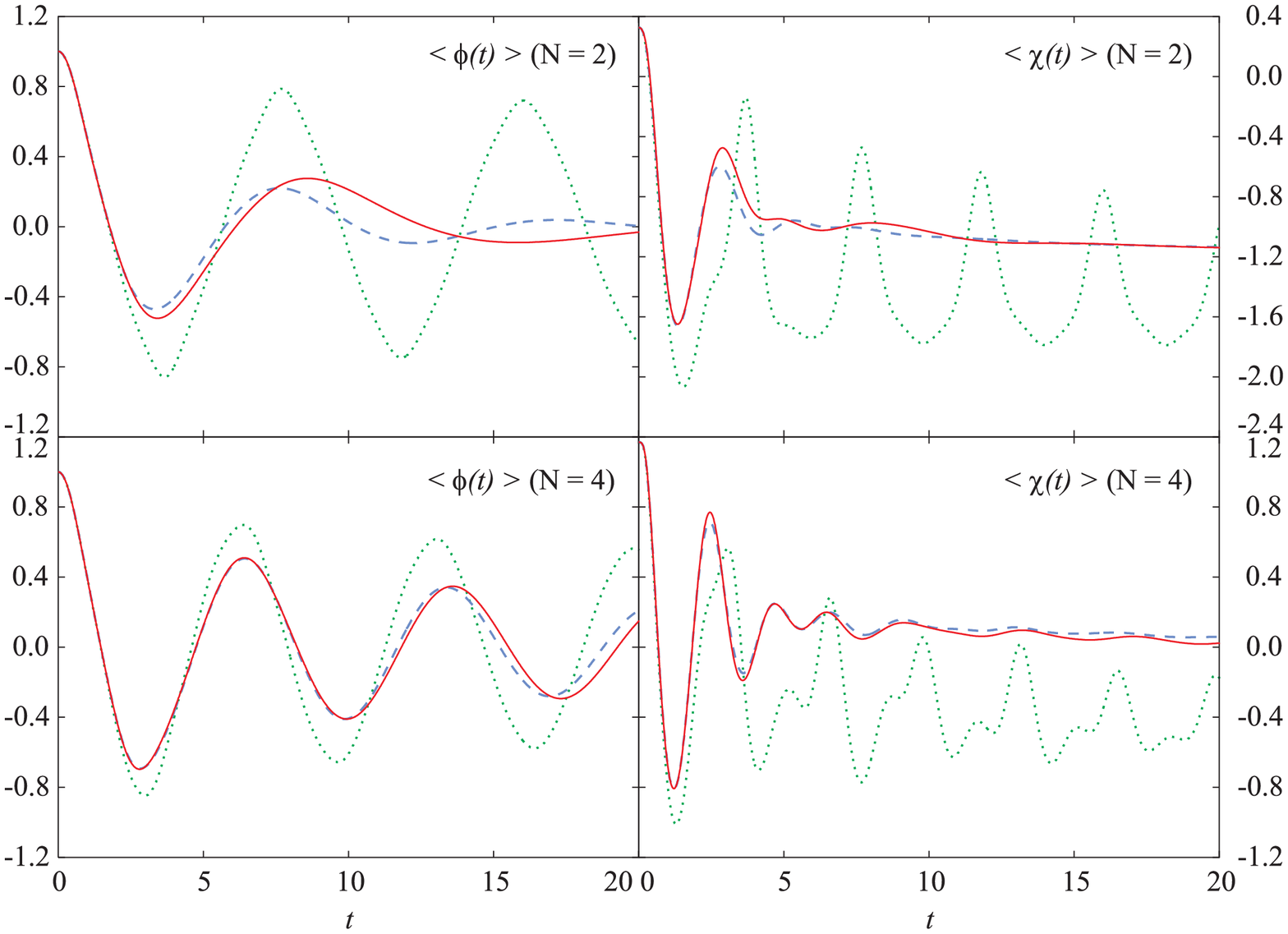}
   \caption{\label{fig:one_field_q} One-field scenario: quantum field
   theory.
   }
\end{figure}

\begin{figure}[h!]
   \includegraphics[height=0.26\textheight]{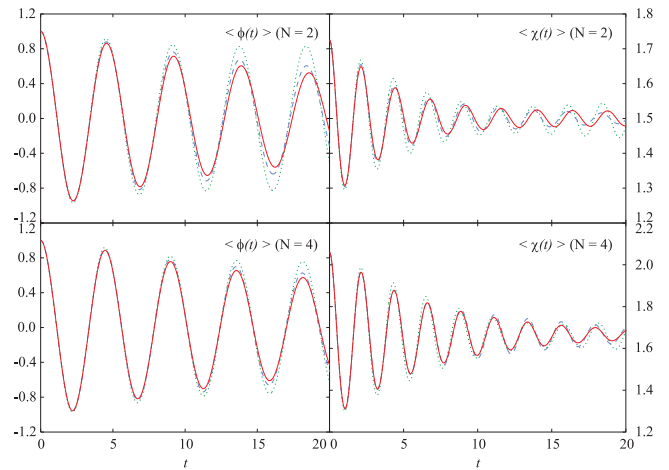}
   \caption{\label{fig:Ncopies_cl} N-copies scenario: classical field
   theory.
   }
\end{figure}

\begin{figure}[h!]
   \includegraphics[height=0.26\textheight]{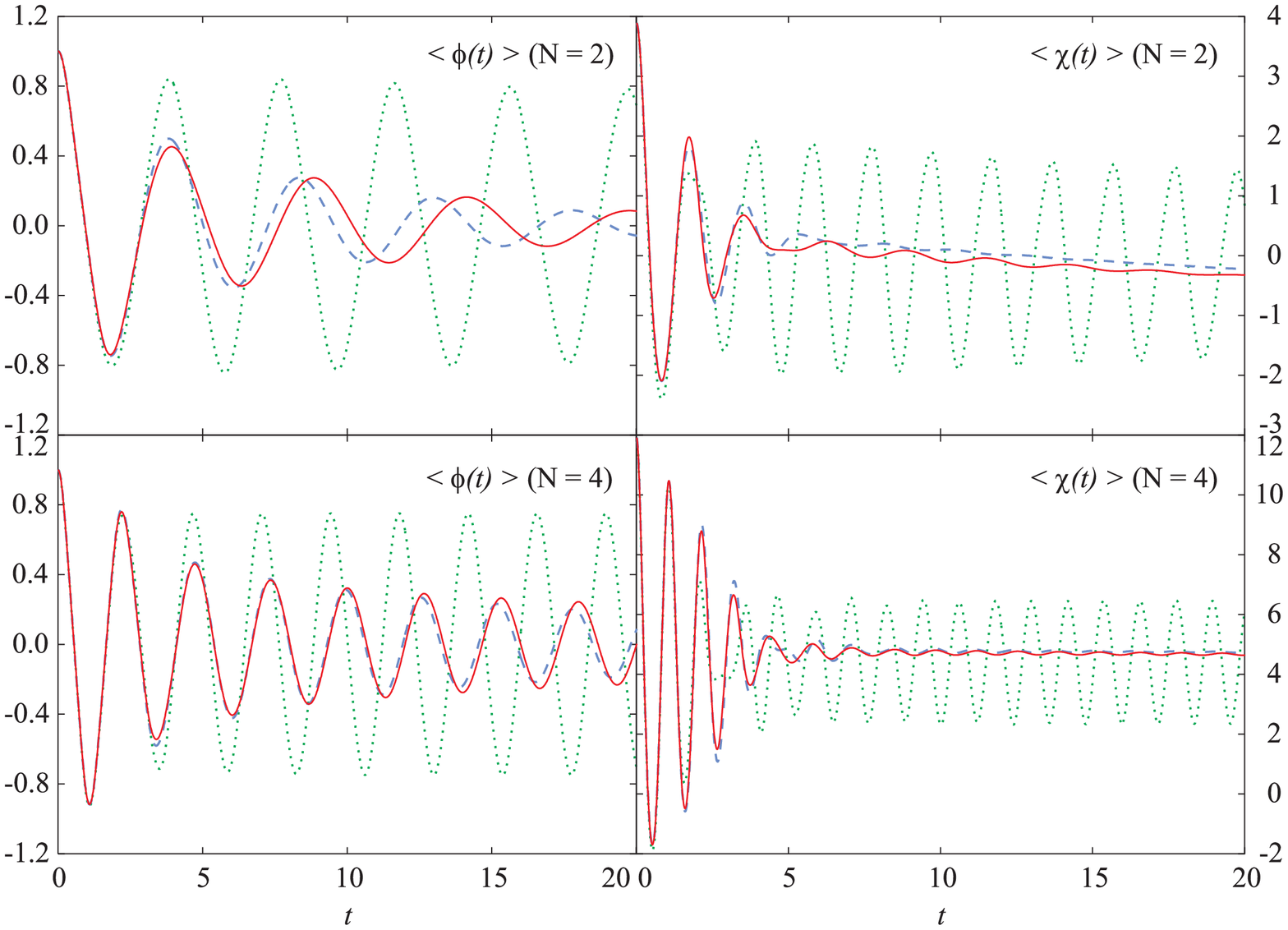}
   \caption{\label{fig:Ncopies_q} N-copies scenario: quantum field
   theory.
   }
\end{figure}

Figures~\ref{fig:one_field_cl} and~\ref{fig:one_field_q} present
the time evolution of the one-point functions for the one-filed
scenario corresponding to the O(2) and O(4) models, for the CFT
and QFT case, respectively. Similar results are shown in
Figs.~\ref{fig:Ncopies_cl} and~\ref{fig:Ncopies_q} for the
N-copies scenario.

One striking feature in reviewing these results, is the fact that
the NLO corrections are more important in the quantum than in the
classical case. The Hartree result represents a poor approximation
in QFT, while in CFT it is arguably doing pretty well. This trend
is still apparent in the $N=10$ results (not shown). For $N=10$,
the differences between 2PI-1/N and BVA for the QFT case have
essentially disappeared, but Hartree is still considerably
different from NLO in the QFT case.

\section{conclusions}
\label{sec:conclusions}

We have shown here first calculations of the O(N) model in
classical and quantum field theory of the BVA and 2PI-1/N for $N >
1$ in 1+1 dimensions. For $N=1$, earlier calculations had shown
substantial differences between these two approximations for the
quantum field theory case.  Our results indicate that these
differences diminish for larger values of $N$. The BVA and 2PI-1/N
approximations differ through contributions which are proportional
to $g^2$. Since $g=\lambda/N$, and we chose to keep the coupling
constant $\lambda$ fixed, in order to approach the limit of the
physically interesting realization of this model (see
Refs.~\cite{r:KDC96,r:KCMPKnp95}), this trend was expected, and we
have in fact observed it earlier in our previous studies of the
quantum mechanical version of this model~\cite{r:MCD01}. What is
gratifying is the fact that already for $N=4$ the two
approximations agree rather well. This gives us hope that our
future calculations of the O(4) linear model in 3+1 dimensions
will provide reliable predictions, converged with respect to the
approximation order. Of course, further numerical evidence will be
obtained by carrying out a similar study by carrying out
simulations in 2+1 and 3+1 dimensions. An additional study will be
pursued in order to provide a better treatment of the three-point
vertex, a study which is clearly necessary in order to reduce the
differences between the exact and BVA results in 1+1 dimensional
classical field theory. It is important to remember that no exact
calculations are available for QFT, unlike the CFT case where
lattice simulations have been performed. Therefore, the fact that
the BVA and 2PI-1/N agree, does not necessarily imply that these
approximations also agree with the exact result, and numerical
evidence supporting the fact that these further corrections do not
affect the results in the O(N) model for $N>1$, is critical to
assert the credibility of the results obtained in this approach.

%
%

\begin{acknowledgments}

Numerical calculations are made possible by grants of time on the
parallel computers of the Mathematics and Computer Science
Division, Argonne National Laboratory. BM would like to
acknowledge useful discussions with John Dawson, Fred Cooper and
J\"urgen Berges.

\end{acknowledgments}

%
%
%
\bibliography{johns}
%
%
\end{document}